\documentclass{iopart}
\usepackage{graphicx}

\usepackage{latexsym}
\usepackage{amssymb}

\begin{document}

\title[]
{Proper Size of the Visible Universe in FRW Metrics with Constant Spacetime Curvature}

\author{Fulvio Melia}
\address{Department of Physics, The Applied Math Program, and Department of Astronomy,\\  
The University of Arizona, AZ 85721, USA}

\begin{abstract}
In this paper, we continue to examine the fundamental basis for the Friedmann-Robertson-Walker (FRW)
metric and its application to cosmology, specifically addressing the question: What is the {\it proper} size of the visible
universe? There are several ways of answering the question of size, though often with an incomplete understanding
of how far light has actually traveled in reaching us today from the most remote sources. The difficulty usually
arises from an inconsistent use of the coordinates, or an over-interpretation of the physical meaning of quantities such 
as the so-called proper distance $R(t)=a(t)r$, written in terms of the (unchanging) co-moving radius $r$ and 
the universal expansion factor $a(t)$. In this paper, we prove for the five non-trivial FRW metrics with constant
spacetime curvature that, when the expansion began from an initial singularity, the visible universe today has a 
proper size equal to $R_{\rm h}(t_0/2)$, i.e., the gravitational horizon at half its current age. The exceptions are 
de Sitter and Lanczos, whose contents had pre-existing positions away from the origin. In so doing, we confirm
earlier results showing the same phenomenon in a broad range of cosmologies, including $\Lambda$CDM, based 
on the numerical integration of null geodesic equations through an FRW metric.
\end{abstract}

\pacno{04.20.Ex, 95.36.+x, 98.80.-k, 98.80.Jk}

\section{Introduction}
Recent efforts aimed at providing a better understanding of the fundamental basis for the 
Friedmann-Robertson-Walker (FRW) metric and its application to cosmology have uncovered 
several previously unrecognized properties relevant to the interpretation of cosmological 
data.  The standard model of cosmology ($\Lambda$CDM) is only marginally consistent with these developing
theoretical considerations, reflected in the growing tension between its predictions and
what is actually observed, both in the cosmic microwave background (CMB) and the unexpected
early appearance of quasars and galaxies at high redshift, and in the matter distribution, gamma-ray
burst rate and Type Ia supernovae in the nearby Universe.

For example, the use of Birkhoff's theorem and its corollary \cite{birkhoff23} has shown that the Universe 
possesses a gravitational horizon (with radius $R_{\rm h}$) coincident with the better known Hubble 
sphere emerging empirically from the observed universal expansion \cite{melia07}. This new insight has 
allowed us to consider the impact of strictly adhering to the requirements of both the 
Cosmological principle and Weyl's postulate \cite{weyl23}, which together force $R_{\rm h}$
to always be equal to $ct$, the distance light could have traveled during a time $t$ since the
big bang \cite{melia12}. $\Lambda$CDM agrees with this constraint only
partially, oddly very early in the universal expansion close to the Planck time, 
and more recently, where the various observations are telling us that $R_{\rm h}(t_0)\approx 
ct_0$ today---but not in between.  For a summary of how the current cosmological data compare with the
condition $R_{\rm h}=ct$ and the predictions of the standard model, see references \cite{spergel03,
copi09,melia12a,melia12b,melia13,meliamaier13}.

This fundamental approach to the study of the cosmological spacetime has also
allowed us to examine the nature of cosmological redshift $z$ in FRW metrics with constant spacetime 
curvature. We recently showed that the interpretation of $z$ as due to the `stretching' of space is 
coordinate dependent \cite{melia12c}. An equally important outcome of this study has been a greatly 
improved understanding of how null geodesics behave in FRW, allowing us to better appreciate which 
sources are actually observable today. We recently confirmed the importance of 
$R_{\rm h}$ in delimiting the size of the observable universe \cite{bikwa12,melia12d}
by proving that in all cosmologies with an equation-of-state parameter $w\ge 
-1$, where the pressure $p$ and density $\rho$ are related by the expression $p=w\rho$, 
no light rays reaching us today could have ever attained a proper distance $R(t)$ greater than
$R_{\rm h}(t_0)$. 

A principal motivation for the present paper is actually another interesting result that 
emerged from the numerical integration of the null geodesics in reference \cite{bikwa12}. 
There, we showed that for a broad range of cosmologies, including $\Lambda$CDM, no null 
geodesics reaching us today (at time $t_0$) could have ever started from, or reached,
a proper distance greater than $\sim ct_0/2$ away from us. Our purpose here is to 
examine the fundamental basis for this constraint, and we will  prove that in FRW 
metrics with a constant spacetime curvature, the most distant sources we see 
today---particularly the CMB---emitted their light at time $(1/2)t_0$ from a proper 
distance $R_{\rm h}(t_0/2)$ away, which therefore defines the size of the visible 
universe today. 

Applied to the CMB, this result may seem paradoxical because the time 
$t_e$ at recombination was presumably much earlier than $(1/2)t_0$. 
Needless to say, this issue has itself caused confusion over the years, with some
workers believing that light must have therefore traveled a proper distance $c(t_0-t_e)$
in reaching us. For example, a recent recalibration (by $\Delta t\sim 2$ Gyr) 
of the age of extragalactic eclipsing binaries was used to stretch the cosmic distance 
ladder by $\sim c\,\Delta t$ \cite{bonanos06}. Similarly, conclusions concerning 
the Universe's topology are often based on how far light has traveled since 
the big bang \cite{cornish03,vaudrevange12}. And an older (often cited)
publication on distance measures makes several incorrect assocations 
between how far light could have traveled and the inferred distance
to horizons \cite{davis04}. But it is easy to demonstrate that the proper distance
to a source is not equal to the light-travel distance, and that the difference is 
merely due to the time dilation between frames moving at relative speeds close 
to $c$. In other words, we shall see that whereas $t_e$ may be close to $0$
(for, say, the CMB), the corresponding time on clocks at rest with respect to us 
was dilated significantly to a value $\sim(1/2)t_0$ ($\gg t_e$).

\section{The FRW Metrics with Constant Spacetime Curvature}
The high degree of symmetry afforded by the FRW metric is a direct consequence
of the Cosmological principle and Weyl's postulate, which together require that
any distance in the cosmos be expressible as the product of an unchanging
comoving radius $r$ and a universal expansion factor $a(t)$ depending only on
the cosmic time $t$ (see \cite{melia12e} for a pedagogical description). The 
Friedmann-Robertson-Walker (FRW) metric for a spatially homogeneous 
and isotropic three-dimensional space may be written in the general form,
\begin{equation}
ds^2 = c^2 dt^2 - a^2(t)[dr^2 (1 - kr^2)^{-1} + r^2(d\theta^2 + \sin^2\theta d\phi^2)]\;,
\end{equation}
where $\theta$ and $\phi$ are the corresponding angular coordinates in 
the comoving frame. The spatial curvature constant $k$ is $+1$ for a closed 
universe, $0$ for a flat, open universe, or $-1$ for an open universe. 

But one must be careful in using this simplification, because the so-called proper 
distance $R(t)=a(t)\int dr\;(1-kr^2)^{-1/2}$ in
these coordinates is not measured using rulers and clocks at rest with respect
to an individual observer; instead, $R(t)$ represents a community distance, compiled 
from the infinitesimal contributions of myriads of observers lined up between the
endpoints, all at the same time $t$ \cite{weinberg72}. Of course, there is nothing intrinsically
wrong with the usage of $R(t)$ as a measure of distance---but only so long as 
one does not over-interpret its physical meaning. For example, a source at $R$
with $dR/dt>c$ is not receding ``superluminally," because although 
$c$ is measured with rulers and clocks at rest with respect to  an individual 
observer, $dR/dt$ is not (we will return to this shortly).

In previous applications \cite{melia12c}, we had demonstrated that a single 
observer can assess the speed of expansion relative to $c$ only in terms of his 
proper distance and proper time, both measured on rulers and clocks at rest with 
respect to himself (see Equations~3 and 4 below).  Only then is the speed 
of light invariant---and always equal to $c$---and a true upper limit to the speed 
of any object in the cosmos. It is in this context, therefore, that to meaningfully 
address the question of how big the visible universe is, the most straightforward 
way is to first find an alternative set of coordinates $dx^\mu$ to rewrite the 
FRW metric in its static form, 
\begin{equation}
ds^2=g_{\mu\nu}\,dx^\mu dx^\nu\;,
\end{equation}
where the metric coefficients $g_{\mu\nu}$ $(\mu,\nu=0,1,2,3)$ are independent
of time $x^0$, because only then can one claim that the distance 
and time are being measured at rest with respect to the observer. 
The form of the metric in Equation~(1) clearly does not satisfy this condition
because $g_{rr}$, $g_{\theta\theta}$ and $g_{\phi\phi}$ are all functions
of $t$ through the expansion factor $a(t)$. The FRW metrics that can be 
transformed in this fashion are those with a constant spacetime curvature 
\cite{melia12c}. Throughout this paper, we will write $x^\mu=(cT,R,\theta,\phi)$ 
for the coordinates that render the FRW metric static. 

As it turns out, there are exactly six such metrics \cite{robertson68}, though
one of these---the Minkowski spacetime---is highly trivial; in each of the five 
remaining cases, a transformation of coordinates permits us to write these solutions 
in static form \cite{florides80}. In the following sections, we will consider 
each of these in turn, the Milne Universe, de Sitter space, anti-de Sitter space,
an open Lanczos-like Universe, and the Lanczos Universe itself.  But we
shall also learn that de Sitter and Lanczos \cite{lanczos24} are quite 
different from the rest because they do not begin their expansion from a 
singularity at time $t=0$. The size of the visible universe for these two cases
is therefore revealingly different from that of all the others. 

It is important to stress as we proceed through this exercise that although the 
spacetime curvature is constant in the cases we consider here, it is generally nonzero. 
This is a crucial point because the result we obtain is not just an artifact of a
cosmology without any spacetime curvature (as in the Milne Universe); it is 
actually independent of what the spacetime curvature happens to be. In other
words, static FRW metrics do not simplify the expansion by eliminating the 
effects of gravity (or dark energy, for that matter). Gravitational effects are 
present even when the FRW metric is time-independent, as is well known from 
the Schwarzschild and Kerr spacetimes.

\section{The ``Earliest" Visible Light}
Our detailed proof will be presented in \S4 below, but before we begin that
treatment, it will be helpful for us to consider the essential elements and ideas of
this procedure using the following simple motivational argument. Quite generally, 
an observer's proper length is the spacelike separation given in terms of 
Equation~(2):
\begin{equation}
d\mathcal{L}\equiv\sqrt{-ds^2}=\sqrt{-g_{ij}\,dx^i\,dx^j}
\end{equation}
where, following convention, the Latin indices $i$ and $j$ run from
1 to 3, representing the spatial coordinates only. Similarly, the proper time 
is the timelike separation
\begin{equation}
d\tau\equiv {1\over c}\sqrt{g_{00}\,dx^0\,dx^0}\;.
\end{equation}
These are the spatial and temporal elements the observer must
use in order to claim that the {\it proper} speed of light $\mathcal{V}_\gamma$ is 
$c$. One can show this trivially by using the null condition in Equation~(2), for then 
\begin{equation}
\sqrt{g_{00}}\,c\,dT=\sqrt{-g_{ij}\,dx^i\,dx^j}\;.
\end{equation}
(This condition implies that we are only considering metrics with zero velocity
shift, but for the cases we include here, this subclass is sufficient. See
reference \cite{smarr78} for further details.)

Thus, defining
\begin{equation}
\mathcal{V}\equiv {d\mathcal{L}\over d\tau}=\sqrt{-{g_{ij}\over g_{00}}{dx^i\over dT}{dx^j\over dT}}\;,
\end{equation}
one can see that the proper speed for light, $\mathcal{V}_\gamma$, is 
always equal to $c$, irrespective of how curved the spacetime happens to be.

The proper speed $\mathcal{V}$ should not be confused with the so-called
``co-ordinate" speed
\begin{equation}
v\equiv\sqrt{-{\eta_{ij}\over \eta_{00}}{dx^i\over dT}{dx^j\over dT}}\;,
\end{equation}
where $\eta_{\alpha\beta}$ is the Minkowski metric tensor. The coordinate
speed can take on any value, even much greater than $c$. However, the
proper speed for particles and objects with mass must always be less than 
$c$ because $ds^2> 0$. That is, since
\begin{equation}
c\,d\tau> d\mathcal{L}\;,
\end{equation}
we must have $\mathcal{V}< c$. For example, in the widely known Schwarzschild
metric for a central mass $M$, we have $g_{00}=(1-2GM/c^2 r)$ and 
$g_{rr}=(1-2GM/c^2r)^{-1}$. Therefore, a static observer will see a photon
approaching the event horizon located at $R_S\equiv 2GM/c^2$ with a proper speed
$\mathcal{V}_\gamma=c$, whereas its coordinate speed $v_\gamma=
\mathcal{V}_\gamma (1-2GM/c^2 r)$ actually goes to zero.

Let us now consider sources of light expanding radially away from the observer
in an FRW spacetime. (A more formal background to the discussion in this section
may be found, e.g., in references \cite{ellis69,smarr78}.)
With reference to Equation~(2) written in static form, the proper time and 
proper distance in this frame are, respectively,
\begin{equation}
d\tau = \sqrt{g_{TT}}\,dT\;,
\end{equation}
and
\begin{equation}
d\mathcal{L}=\sqrt{-g_{RR}}\,dR\;.
\end{equation}
Thus, the proper speed of light in this frame is
\begin{equation}
\mathcal{V}_\gamma= \sqrt{-{g_{RR}\over g_{TT}}}{dR_\gamma\over dT}\;.
\end{equation}
Reference to ``superluminal" motion in cosmology is often based on the 
coordinate speed, $v_\gamma\equiv dR_\gamma/dT=c\sqrt{-g_{TT}/g_{RR}}$ which, 
as we have said, is not the speed of light measured on rulers and clocks at rest 
with respect to the observer. This coordinate speed diverges for sources
beyond the Hubble radius.

We can now ask the question ``What was the earliest time $T_e$ at which light 
we are receiving right now could have been emitted?" As shown in reference
\cite{bikwa12},
the earliest {\it cosmic} time $t_e$ at which this light could have been emitted
was 0, because all null geodesics linking us to our past actually started at
the origin of the coordinates for $t\rightarrow 0$. But this is not true
of the time $T_e$ on the individual observer's clocks, because when 
viewed in terms of the observer's coordinates, the sources were not at 
the origin when they emitted the light we see today.

Instead, the sources had to first travel out to a proper distance equal
to that traversed by light once it was emitted back towards us. Designating
the proper speed of a source as $\mathcal{V}_S$, we therefore see that
\begin{equation}
\int_0^{T_e}\mathcal{V}_S(\sqrt{g_{TT}}\,dT)=\int_{T_e}^{T_0}
\mathcal{V}_\gamma(\sqrt{g_{TT}}\,dT)\;,
\end{equation}
where $T_0$ is the present time. But since the proper lightspeed
is always $c$, we can write the somewhat simpler expression
\begin{equation}
\int_0^{T_e}\mathcal{V}_S(\sqrt{g_{TT}}\,dT)=\int_{T_e}^{T_0}
c(\sqrt{g_{TT}}\,dT)\;.
\end{equation}
However, we have just argued that the maximum proper speed of any
particle or object is $\mathcal{V}_S\rightarrow c$. The earliest
time $T_e$ that a source could have emitted light just reaching
us today corresponds to the fastest among them---those that
reached the greatest proper distance in the shortest time.
In other words, the earliest time $T_e$ corresponds to
$\mathcal{V}_S=c$, and therefore the condition on
$T_e$ becomes
\begin{equation}
\int_0^{T_e}\sqrt{g_{TT}}\,dT\rightarrow\int_{T_e}^{T_0}\sqrt{g_{TT}}\,dT\;,
\end{equation}
which has the obvious solution $T_e=(1/2)T_0$. But
$T_0=t_0$, since $t$ is actually the proper time on the clock at rest
with respect to us at our location, and we thus infer that the earliest 
time light visible to the observer could have been emitted must have 
been $T_e=(1/2)t_0$. In the next section, we will prove this result 
for each of the FRW metrics with constant spacetime curvature that
begin their expansion from an initial singularity, and then we will discuss 
what it means to have $T_e>>t_e$ (i.e., $T_e>>0$).

\section{Detailed Proof}

\subsection{The Milne Universe}
The Milne universe \cite{milne33} has no density ($\rho=0$) and is characterized
by a spatial curvature $k=-1$. It therefore corresponds to a simple
solution of Einstein's equations with
\begin{equation}
a(t)=ct\;,
\end{equation}
in which the scale factor grows linearly with time at a rate equal to
the speed of light $c$. Since the acceleration $\ddot{a}(t)$ is
zero in this cosmology, one might expect such a universe to
have zero spacetime curvature and be a mere 
re-parametrization of Minkowski space. Indeed,
Milne built this type of expansion based solely on special relativity,
without any constraints imposed by the more general theory.

To cast the FRW metric for the Milne Universe in its static form 
\cite{abramowicz07,cook09}, we 
first introduce the co-moving distance variable $\chi$, defined in
terms of $r$ according to
\begin{equation}
r=\sinh\chi\;,
\end{equation}
which allows us to write
\begin{equation}
ds^2=c^2dt^2-c^2t^2[d\chi^2+\sinh^2\chi\,d\Omega^2]
\end{equation}
where, for simplicity, we have also introduced the notation
$d\Omega^2\equiv d\theta^2+\sin^2\theta\,d\phi^2$.
The transformation that brings Equation~(10) into a static
form is
\begin{eqnarray}
T&=&t\cosh\chi\nonumber\\
R&=&ct\sinh\chi\;,
\end{eqnarray}
for then
\begin{equation}
ds^2=c^2dT^2-dR^2-R^2d\Omega^2\;.
\end{equation}
Clearly, a null geodesic in this frame has
\begin{equation}
c\,dT=-dR
\end{equation}
(for light approaching us along a radius), and therefore
\begin{equation}
R_e=c(T_0-T_e)\;,
\end{equation}
where the proper distance $\mathcal{L}_e$ to the source in the static coordinate 
system is here identical to $R_e$ because $g_{TT}=g_{RR}=1$, and $T_e$ and 
$T_0$ are the emission and observation times, respectively, in this frame.
To be clear, since the metric coefficients $g_{\mu\nu}$ in Equation~(12)
are independent of $T$, the quantities defined in Equation~(11)
are now the coordinates for an individual observer, because time $T$
is being measured on clocks at rest in his frame.

Suppose now that at cosmic time $t_e$, a source at $\chi_e$ emits the light
reaching the observer today. The corresponding proper time $T_e$ on the 
observer's clock at the location of the source will be greater than $t_e$ due 
to the effects of relativistic time dilation. The most distant sources were 
moving at proper speeds close to $c$ in this frame when they emitted 
their light at $t_e$ \cite{melia12c}, so the time dilation between $T_e$ 
and $t_e$ for them approaches infinity. For this exercise, we may even 
allow $t_e\rightarrow 0$, in order to examine the time at which the CMB 
was produced. From Equation~(11), we see that when $t_e\rightarrow 0$,
\begin{equation}
{T_e\over t_e}=\cosh\chi_e\gg 1\;.
\end{equation}
In this limit, therefore, $\chi_e\rightarrow\infty$, and so we may write
\begin{equation}
{T_e\over t_e}\rightarrow {1\over 2}e^{\chi_e}\;.
\end{equation}
But we also know from Equation~(10) that a null geodesic satisfies
\begin{equation}
\int_0^{\chi_e}d\chi'=\int_{t_e}^{t_0}{dt'\over t'}\;,
\end{equation}
and so
\begin{equation}
{t_0\over t_e}=e^{\chi_e}\;.
\end{equation}
A comparison between Equations~(16) and (18) immediately tells us that,
in the limit $t_e\rightarrow 0$,
\begin{equation}
T_e\rightarrow {1\over 2}t_0\;,
\end{equation}
a simple and beautiful result that confirms our supposition from the previous
section, and one that will be repeated with each subsequent cosmology we 
consider (except for de Sitter and Lanczos, as we have already anticipated). 

Thus, in the Milne Universe, the greatest proper distance (defined in Equation~14) traveled 
by light reaching the observer today is only $(1/2)ct_0$, though the sources there 
look very young due to the effects of time dilation between the source frame and that of
the observer. In other words, though the CMB may have been produced at cosmic time
$t_e\rightarrow 0$ in the co-moving frame, that event occurred at time $T_e=(1/2)t_0$
in the observer's frame, and though the light signal carries information pertaining to 
those earliest moments in the Universe's expansion history, according to Equation~(14)
it has therefore only traveled a proper distance $(1/2)ct_0$ in reaching the observer. 

Later in the discussion 
section we will describe in greater detail what is actually happening, but the basic principle 
is rather simple---since in Milne the expansion began from a singularity at time 
$t=0$, it took the most distant sources a time $(1/2)t_0$ to reach a proper distance 
($R_e=ct_0/2$) from which the most distant light signal could then have reached 
the observer by time $t_0$. This feature is illustrated schematically in Figure~1.
Note that formally, $R_{\rm h}=c/H$, and since here $H\equiv \dot{a}/a=1/t$,
we may also write this result in the form $R_e=R_{\rm h}(t_0/2)$. Observationally,
we recognize $R_{\rm h}$ as the Hubble radius.

\begin{figure} 
\begin{center}
\includegraphics[scale=0.5]{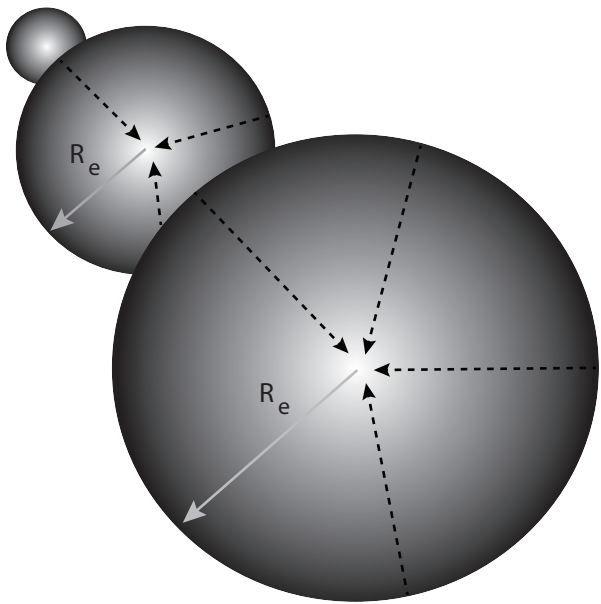}
\end{center}
\caption{Illustration of the maximum proper distance light could have traveled along (dashed) null geodesics
in reaching us as the Universe expands through a metric with constant spacetime curvature. At time $t$ in 
these spacetimes, we see light only from sources that produced their signal after $(1/2)t$, and were located 
within a sphere of maximum extent $R_e$. In all cosmologies with constant spacetime curvature beginning 
their expansion from an initial singularity, including the Milne universe, $R_e=R_{\rm h}(t_0/2)$, the 
gravitational horizon at half the time $t_0$ elapsed since the big bang.}
\end{figure}

\subsection{Anti de Sitter Space}
A Universe with negative mass density and spatial curvature $k=-1$ is known as anti-de 
Sitter space, due to its negative spacetime curvature. This metric is given by
\begin{equation}
ds^2=c^2dt^2-\left({cb}\right)^2\sin^2\left(t/b\right)\left[{dr^2\over 1+r^2}+
r^2d\Omega^2\right]\;,
\end{equation}
where clearly the expansion factor is now $a(t)={cb}\sin\left(t/b\right)$.
The coordinate transformation
\begin{equation}
R={cbr}\sin(t/b)\;,
\end{equation}
and
\begin{equation}
\tan(T/b)=\left(1+r^2\right)^{1/2}\tan(t/b)\;,
\end{equation}
produces the static form of the metric,
\begin{equation}
ds^2=\left[1+\left({R\over cb}\right)^2\right]c^2dT^2-
\left[1+\left({R\over cb}\right)^2\right]^{-1}dR^2-R^2d\Omega^2\;.
\end{equation}

Now, along a geodesic connecting the emission point $r_e$ at time $t_e$ with the
observer at $r=0$ and time $t_0$, 
\begin{equation}
\int_0^{r_e}{dr\over\sqrt{1+r^2}}=\int_{t_e/b}^{t_0/b}{du\over \sin(u)}\;,
\end{equation}
which has the solution (Melia 2012c)
\begin{equation}
r_e={1\over 2}\left({\tan(t_0/2b)\over \tan(t_e/2b)}
-{\tan(t_e/2b)\over \tan(t_0/2b)}\right)\;.
\end{equation}
Thus, for a source emitting light very early in the Universe's history, i.e., for
$t_e\rightarrow 0$,
\begin{equation}
r_e\rightarrow {1\over 2}{\tan(t_0/2b)\over t_e/2b}\;.
\end{equation}
In this limit, $r_e$ clearly diverges, so the factor $\sqrt{1+r_e^2}$ in
the equation for $T_e$,
\begin{equation}
\tan(T_e/b)=(1+r_e^2)^{1/2}\tan(t_e/b)\;,
\end{equation}
simply becomes $r_e$. In addition, $\tan(t_e/b)$ in this limit reduces to
$t_e/b$, and so
\begin{equation}
\tan(T_e/b)\rightarrow {1\over 2}{\tan(t_0/2b)\over t_e/2b}{t_e\over b}=\tan(t_0/2b)\;,
\end{equation}
which again leads to the beautiful result that
\begin{equation}
T_e\rightarrow {1\over 2}t_0\;.
\end{equation}
This is the observer's coordinate time at which the most distant sources visible to him
at $t_0$ produced their light in anti de Sitter space.

\subsection{A Lanczos-like Universe with $k=-1$}
In co-moving coordinates, the third FRW metric with constant spacetime curvature (and
$k=-1$) may be written \cite{robertson68,florides80}
\begin{equation}
ds^2=c^2dt^2-\left({cb}\right)^2\sinh^2(t/b)\left[{dr^2\over 1+r^2}+
r^2d\Omega^2\right]\;,
\end{equation}
where $a(t)=(cb)\sinh(t/b)$, and $b$ is a constant (though clearly not the Hubble constant
$H\equiv \dot{a}/a$). The metric may also be written in static form with the transformation
\begin{equation}
R={cbr}\sinh(t/b)\;,
\end{equation}
and
\begin{equation}
\tanh(T/b)=\left(1+r^2\right)^{1/2}\tanh(t/b)\;,
\end{equation}
which together allow us to write the interval in the form
\begin{equation}
ds^2=\left[1-\left({R\over cb}\right)^2\right]c^2dT^2-
\left[1-\left({R\over cb}\right)^2\right]^{-1}dR^2-R^2d\Omega^2\;,
\end{equation}
identical (in terms of $R$ and $T$) to the actual Lanczos metric we shall 
examine below. 

As before, we calculate the co-moving radius $r_e$ to a source at time $t_e$
using the geodesic equation
\begin{equation}
\int_0^{r_e}{dr\over\sqrt{1+r^2}}=\int_{t_e/b}^{t_0/b}{du\over \sinh(u)}\;,
\end{equation}
whose solution is
\begin{equation}
\sinh^{-1}\left(r_e\right)=\ln\left(\tanh\left[t_0/2b\right]\right)-\ln\left(\tanh\left[t_e/2b\right]\right)\;.
\end{equation}
That is,
\begin{equation}
r_e={1\over 2}\left({\tanh(t_0/2b)\over \tanh(t_e/2b)}
-{\tanh(t_e/2b)\over \tanh(t_0/2b)}\right)\;.
\end{equation}

The most distant sources visible by the observer at time $t_0$ emitted their light
at time $t_e\rightarrow 0$, so their comoving radius is evidently
\begin{equation}
r_e\rightarrow {1\over 2}{\tanh(t_0/2b)\over t_e/2b}\;.
\end{equation}
As was true for anti de Sitter space, this radius diverges for the earliest times,
and therefore $\sqrt{1+r_e^2}\rightarrow r_e$. Thus,
\begin{equation}
\tanh(T_e/b)\rightarrow {1\over 2}{\tanh(t_0/2b)\over t_e/2b}{t_e\over b}=\tanh(t_0/2b)\;,
\end{equation}
so once again
\begin{equation}
T_e\rightarrow {1\over 2}t_0\;.
\end{equation}

\subsection{de Sitter Space}
We next consider the situation with de Sitter which, unlike Milne,
is a cosmology in which objects not only recede from each other,
but also accelerate under the influence of gravity. However, unlike
anti de Sitter space and the open Lanczos-like universe we have just
considered, de Sitter space does not begin its expansion from a
singularity.

The de Sitter cosmology \cite{desitter17} corresponds to a universe devoid
of matter and radiation, but filled with a cosmological constant whose
principal property is the equation of state $p=-\rho$. The FRW metric in
this case may be written
\begin{equation}
ds^2 = c^2 dt^2-e^{2Ht} [dr^2 + r^2d\Omega^2]\;,
\end{equation}
where $k = 0$ and the expansion factor has the specific form
\begin{equation}
a(t)=e^{Ht}\;,
\end{equation}
in terms of the Hubble constant $H$.

Unlike the Milne model, de Sitter space contains mass-energy
(in the form of a cosmological constant). The transformation of
coordinates that brings the metric (Equation~40) into its stationary
form is as follows:
\begin{eqnarray}
R&=&a(t)r\nonumber\\
T&=&t-{1\over 2H}\ln\Phi\;,
\end{eqnarray}
where
\begin{equation}
\Phi\equiv 1-\left({R\over R_{\rm h}}\right)^2\;,
\end{equation}
and
\begin{equation}
R_{\rm h}\equiv {c\over H}
\end{equation}
is the gravitational (or Hubble) radius. With these, the de Sitter
metric becomes
\begin{equation}
ds^2=c^2\Phi\,dT^2-\Phi^{-1}dR^2 - R^2d\Omega^2\;,
\end{equation}
and notice that all the metric coefficients are now independent
of time $T$. Of course, the chief difference between this case and
that exhibited in Equation~(12), is that the de Sitter spacetime
is curved, and therefore both $g_{TT}$ and $g_{RR}$ depend on
the spatial coordinates \cite{melia07}. The form of the metric in
Equation~(45) is how de Sitter himself first presented his now
famous solution.

To be precise, de Sitter does not actually have an initial singularity because
$a(t)\rightarrow 1$ as $t\rightarrow 0$. Rather, the transformation
exhibited in Equations~(42)--(44) demonstrates the dependence of
$R$ and $T$ only on the time difference between $t$ and an
initial time $t_i$. The form of these equations corresponds to
the choice $t_i=0$. In talking about the proper size of the visible
universe in de Sitter (see \S~V), we therefore necessarily refer to how
far sources and light have moved during a time $t_0-t_i$.
Again, in the expressions that follow, we will adhere to the
convention that $t_i=0$.

As we did with Milne, let us now consider the time $T_e$ at 
which the source located at $R_e(t_e)$ emits the light we see 
today, corresponding to the proper time $t_e$ in its own rest frame. 
It will be useful for us to borrow a result we derived earlier \cite{melia12c},
allowing us to express the redshift in the form
\begin{equation}
1+z={1\over 1-{R_e(t_e)/R_{\rm h}}}\;.
\end{equation}
This follows very easily from the well-known formulation \cite{weinberg72}
\begin{equation}
1+z={a(t_0)\over a(t_e)}\;,
\end{equation}
or
\begin{equation}
1+z=\exp\left[H(t_0-t_e)\right]\;.
\end{equation}
According to Equations~(42) and (43),
\begin{equation}
T_e=t_e-{1\over 2H}\ln\left[1-\left({R_e(t_e)\over R_{\rm h}}\right)^2\right]\;,
\end{equation}
so that
\begin{equation}
T_e=t_e-{1\over 2H}\ln\left[1-\left({R_e(t_e)\over R_{\rm h}}\right)\right]
-{1\over 2H}\ln\left[1+\left({R_e(t_e)\over R_{\rm h}}\right)\right]\;.
\end{equation}
This equation may be manipulated further, yielding
\begin{equation}
T_e=t_e-{1\over 2H}\ln e^{-H(t_0-t_e)}
-{1\over 2H}\ln\left({1+2z\over 1+z}\right)\;.
\end{equation}
And this leads to the result we were seeking,
\begin{equation}
T_e={1\over 2}t_0+{1\over 2}t_e-{1\over 2H}\ln\left({1+2z\over 1+z}\right)\;.
\end{equation}

To find the time $T_e$ in the observer's frame corresponding to the 
earliest emission of observable light in the universe, we put $t_e\rightarrow 0$
and $z\rightarrow\infty$, so that
\begin{equation}
T_e\rightarrow {1\over 2}t_0-{1\over 2H}\ln 2\;.
\end{equation}
Therefore, when $t_0\gg 1/H$, we obtain the rather remarkable result
that, even in this kind of curved spacetime,
\begin{equation}
T_e\rightarrow {1\over 2}t_0\;.
\end{equation}
This is what one would expect on the basis of our discussion in the previous
section. As we shall see shortly, however, the proper distance associated with
this emission time is not the same as that for the previous three cases.

\subsection{The Lanczos (Closed) Universe}
The fifth, and final, FRW metric with constant spacetime curvature (other than Minkowski)
 is known as the Lanczos Universe, described by the metric
\begin{equation}
ds^2=c^2dt^2-\left({cb}\right)^2\cosh^2(t/b)\left[{dr^2\over 1-r^2}+
r^2d\Omega^2\right]\;,
\end{equation}
where $k=+1$, and the expansion factor is now $a(t)=(cb)\cosh(t/b)$. 
We use the following transformation to render this metric in static form:
\begin{equation}
R={cbr}\cosh(t/b)\;,
\end{equation}
and
\begin{equation}
\tanh(T/b)=\left(1-r^2\right)^{-1/2}\tanh(t/b)\;,
\end{equation}
which together allow us to write the interval as
\begin{equation}
ds^2=\left[1-\left({R\over cb}\right)^2\right]c^2dT^2-
\left[1-\left({R\over cb}\right)^2\right]^{-1}dR^2-R^2d\Omega^2\;.
\end{equation}

But it doesn't take much to realize that this universe is quite different from the others.
For one thing, it's closed ($k=+1$); the others are all open, with $k=0$ or $k=-1$. Moreover,
as in de Sitter, the expansion factor $a(t)$ does not vanish at $t=0$. Instead, $a\rightarrow 
cb$, so this universe does not begin its expansion from a singularity, and because the sources
all occupied pre-existing positions at $t=0$, one should therefore expect that
an observer can detect the light they emitted all the way back to $T_e=0$,
which we now demonstrate formally.

As shown in reference \cite{melia12c}, the co-moving distance to a source emitting light at
$t_e$ is here given by the expression
\begin{eqnarray}
r_e&=&2\left(e^{t_0/b}-e^{t_e/b}\right)\left(1+e^{(t_0+t_e)/b}\right)\times\cr
&\null&\qquad\qquad\qquad\qquad\left(1+e^{2t_0/b}\right)^{-1}\left(1+e^{2t_e/b}\right)^{-1}\;,
\end{eqnarray}
and it is not difficult to show that as $t_e\rightarrow 0$, 
\begin{equation}
r_e\rightarrow \tanh(t_0/b)\;.
\end{equation}
Therefore as $t_e\rightarrow 0$,
\begin{equation}
\tanh(T_e/b)\rightarrow \cosh(t_0/b)\tanh(t_e/b)\rightarrow 0\;,
\end{equation}
for which $T_e\rightarrow 0$. 

\section{Proper Size of an FRW Universe}
In all the cases we have considered, the coordinate distance $R$ in the frame
where all the metric coefficients are independent of time actually coincides 
with the definition of proper radius $R=a(t)r$ in the co-moving
frame.  We may therefore determine the ``proper" size of the visible universe by 
merely calculating $R_e(T_e)$ at the time of emission $T_e$.

The Milne universe is unique among the other FRW metrics because it has
zero spacetime curvature. Therefore $g_{RR}=g_{TT}=1$, and $R$ is exactly equal
to the light-travel distance $cT$, as indicated in Equation~(14). Thus, using Equation~(19),
we see that for Milne, $R_e=(1/2)ct_0=R_{\rm h}(t_0/2)$.

For anti de Sitter space, 
\begin{equation}
R_e=cbr_e\sin(t_e/b)\;,
\end{equation}
and introducing the limiting form of $r_e$ from Equation~(26), we find that
\begin{equation}
R_e\rightarrow cb\tan(t_0/2b)\;,
\end{equation}
which therefore gives, again, $R_e=R_{\rm h}(t_0/2)$.

A similar result follows for the third universe we considered, since in this case
\begin{equation}
R_e=cbr_e\sinh(t_e/b)\;,
\end{equation}
and therefore using the limiting form of $r_e$ from Equation~(37), we find that
\begin{equation}
R_e\rightarrow cb\tanh(t_0/2b)\;.
\end{equation}
Since in this case $H(t)=1/b\tanh(t/b)$, we find that $R_e=R_{\rm h}(t_0/2)$,
as was the case for Milne and anti de Sitter.

These three independent cases all demonstrate the prinicpal result of this
paper---that in a FRW metric with constant spacetime curvature 
expanding from a singularity, the earliest
signal we can see today was produced when the Universe was half its current
age, $t_0$ (as measured on clocks at rest with respect to us), from a proper distance
equal to the size of the gravitational horizon at that time.  As such, this formal derivation
is fully consistent with the numerical calculations reported in reference \cite{bikwa12}, and helps
to explain the conclusions in that paper---that light reaching us today, 
even in the case of $\Lambda$CDM, never
attained a proper distance greater than $R_{\rm h}(t_0/2)$.

Let us now compare this fundamental result with de Sitter and Lanczos, for
which $a(0){\not=0}$. It is straightfoward to see from Equation~(42) that
\begin{equation}
-t_0H=\ln\left(1-\left[{R_e\over R_{\rm h}}\right]^2\right)\;,
\end{equation}
and therefore $R_e\rightarrow R_{\rm h}$ for $t_0H>>1$. Even
though $T_e=t_0/2$, the proper size of the visible de Sitter universe
is nonetheless $R_{\rm h}$, which is a constant. The fact that $T_e{\not=}0$
is entirely due to our choice of following the expansion from a defined initial 
time $t_i=0$. But the structure of this universe is independent of time
because its expansion is eternally exponentiated. So the most distant
sources we can see are always at the gravitational horizon.

There are strong similarities between de Sitter and the final cosmology
we have considered---the Lanczos universe---though the gravitational
horizon in this case is not constant. We see from Equation~(56) that
\begin{equation}
R_e=cbr_e\cosh(t_e/b)\;,
\end{equation}
so evidently 
\begin{equation}
R_e(t_e)\rightarrow cb\tanh(t_0/b)={c\over H(t_0)}=R_{\rm h}(t_0)\;.
\end{equation}
This is a very interesting and important result in itself, because it demonstrates
that, regardless of whether or not the gravitational horizon is moving, the most
distant sources we see today in a universe without an intial singularity coincide
with the location of this horizon today.

Furthermore, notice that even though $R_{\rm h}$ is here not constant, we find that $R_e
\rightarrow cb$ for all $t_0/b\gg 1$, which mirrors the situation with
de Sitter. The principal difference between these last two cosmologies 
and all the others,  is that whereas the sources in the previous
cases first had to travel a proper distance $R_{\rm h}(t_0/2)$ to reach the edge of
the visible universe, the most distant sources in de Sitter and Lanczos
were already situated at $R_e=R_{\rm h}(\infty)$ from the very beginning, 
and therefore the observer can see their light emitted from that maximal distance 
at arbitratily early times. 
 
\section{Discussion and Conclusions}
To fully understand and appreciate the results we have presented in this paper,
one must acknowledge the critical role played by the choice of coordinates in
describing the expansion of the Universe. We had already seen an example
of this, based on how the choice of frames impacts our interpretation of the cosmological 
redshift $z$ \cite{melia12c}. We proved earlier that, although $z$ is conventionally
calculated directly from the expansion factor $a(t)$,  its origin cannot be attributed 
to an expansion of space when viewed in terms of the FRW metric written in 
stationary form. We found that $z$ is actually the cosmological version of
a lapse function encountered more typically in the context of the Schwarzschild
and Kerr metrics. That is, $z$ is simply due to the combined effects of the kinematic
expansion and the gravitational acceleration---but only in terms of the proper
velocity, calculated using the proper distance and proper time for an individual
observer.

In this paper, we have expanded our study of the fundamental aspects of the
cosmic spacetime by using these alternative sets of coordinates to address another
issue that sometimes gives rise to confusion and ambiguity: what is the true size 
of the visible universe?  The question itself is fraught with ambiguity because
it goes without saying that to measure a size, one must have a precise 
definition of distance. General relativity is founded on the basic
principle that $c$ is invariant and is measured to have the same value for
all observers. But what is often overlooked or forgotten is that in order to
make the measurements consistent with this tenet, distances and times
must be determined with devices at rest with respect to the observer.
Only then can he claim that $c$ is an upper limit to all speeds and that
light travels at speed $c$ under all circumstances and at all times.

These notions are particularly important to the question we have addressed
in this paper, especially for cosmologies that begin their expansion from
a singularity at time $t=0$. The reason for this is rather straightforward.
In these cosmologies, all the worldlines of sources we see today started
from the same location---very near the same co-moving point we 
ourselves are now occupying. Clearly, to suggest that 
the light they emitted has traveled a distance $c(t_0-t_e)
\rightarrow ct_0$ since the big bang is quite non-sensical. The
correct statement is that the most distant sources we see today are precisely
those moving at close to proper lightspeed, which reached a 
proper distance $R_{\rm h}(t_0/2)$ before emitting the light 
that is just now reaching us at time $t_0$.

It is remarkable---though obvious in retrospect---how elegantly and beautifully
this simple result emerges from the properties of the FRW metric itself
written  in stationary form, when we take the limit $t_e\rightarrow 0$
for the time at which the light from the most distant sources was emitted.
One of the principal results of our analysis has been the demonstration that
even though $t_e\approx 0$ for these sources, the time measured on
our clocks at rest with respect to us was actually $T_e=(1/2)t_0$. And
now we understand that this effect is entirely due to the time dilation
between us and sources receding at proper speed $c$ when they emitted 
this light.

These conclusions do come with a caveat, however, because most 
of these results are based on the use of FRW metrics with a constant
spacetime curvature, allowing us to find an alternative set of
coordinates to write them in stationary form. Without this option,
we would not yet know how to evaluate distances and times in such
a way as to demonstrate without any doubt how far sources 
could have traveled before emitting the light we see today.
One ought to expect the proper size of the visible universe
to be measurable against the gravitational horizon even
in cases where the spacetime curvature is not constant,
but only future work can establish this result conclusively.

\section*{Acknowledgements}

This research was partially supported by ONR grant N00014-09-C-0032
at the University of Arizona, and by a Miegunyah Fellowship at the
University of Melbourne. I am particularly grateful to Amherst College
for its support through a John Woodruff Simpson Lectureship.  I
am also grateful to the anomymous referees for suggesting improvements
to the manuscript.

\end{document}